\documentclass[12pt]{article}%
\usepackage{amsmath}
\usepackage{amsfonts}
\usepackage{amssymb}
\usepackage{graphicx}%
\providecommand{\U}[1]{\protect\rule{.1in}{.1in}}
\begin{document}
\begin{titlepage}
\begin{flushright}
PUPT-2320\\
hep-th/yymmnnn
\end{flushright}
\vspace{7 mm}
\begin{center}
\huge{Decay of Vacuum Energy}
\end{center}
\vspace{10 mm}
\begin{center}
{\large
A.M.~Polyakov\\
}
\vspace{3mm}
Joseph Henry Laboratories\\
Princeton University\\
Princeton, New Jersey 08544
\end{center}
\vspace{7mm}
\begin{center}
{\large Abstract}
\end{center}
\noindent
This paper studies interacting massive particles on the de Sitter
background. It is found that  the vacuum acts as an inversely populated medium which is able to generate  stimulated radiation. Without back reaction (not considered in this paper) this effect leads to the explosion. It is expected that the proposed  "cosmic laser" mechanism depletes the curvature and  may help to solve the cosmological constant problem.
\vspace{7mm}
\begin{flushleft}
December 2009
\end{flushleft}
\end{titlepage}

\section{\bigskip Introduction}

Dark energy, like the black body radiation 150 years ago, hides secrets of
fundamental physics. An analyses of the de Sitter space is obviously an
important theoretical step towards understanding of this mysterious quantity..

Recently I have argued \cite{pol08} \cite{talk} that this space (considered as
a fixed background) is unstable whenever any interacting field theory is
present. The purpose of this article is to clarify and generalize this
statement, as well as to provide some solvable examples of this instability.
It will be argued that the explosive particle production in the de Sitter
space must continue until the back-reaction becomes relevant and hopefully
will eventually wipe out all traces of the cosmological constant (or positive
curvature or cosmic acceleration). Nature seems to abhor positive curvature
and is trying to get rid of it as fast as it can .

Since the important pioneering works \cite{chern67}, \cite{nacht67}, there has
been a huge and sometimes confusing literature on the subject (see
\cite{anton03} for a review). The reason for the confusion ( apart from the
general difficulty of the problem) often lies in an unappreciated fact that
different setups in the de Sitter space may lead to the different conclusions.
In particular, there is a difference between the global dS space and its
various patches glued to something else.

Let us begin with remembering the reasons for stability of the Minkowski
space. According to Wigner, elementary particles in this space are associated
with the unitary irreducible representations of the Poincare group. This group
acts on a scalar field as $\varphi(x)\rightarrow\varphi(\Lambda x+a),$ where
$\Lambda$ is a Lorentz transformation and $a$ is a translation. At this stage
the representation is still reducible and one has to impose as many invariant
constraints as possible. For the scalar field they are provided by the wave
equation ( $\partial^{2}-m^{2})\varphi=0.$ This justifies Wigner's claim -
solutions of the wave equations correspond to the irreducible representations.

For the real masses these representations have a lowest weight, following from
the fact that the sign of energy is invariant and thus $P_{0}>0.$ The vacuum
is stable because due to the energy conservation it can't create particles. In
terms of the representations it means that the product of these lowest weight
representations doesn't contain a unit one ( describing the vacuum ).

The situation is different for $m^{2}<0$ in which case there is no lowest
weight and thus the tachyons can be generated from the vacuum until being
stopped by their interaction.

In the case of the de Sitter space the group-theoretic situation is quite
different. It has been shown by Wigner that the de Sitter group, $SO(4,1)$ has
the feature he called " ambivalence". Namely, for any group element $g,$ there
is a similarity transformation which takes it to its inverse, $S^{-1}%
gS=g^{-1}$ . From this it follows that all symmetry generators have the
spectrum symmetric under reflection, and thus no lowest weight representation
can exist \footnote{I thank Pierre Ramond for providing me with an unpublished
(?) manuscript by Wigner and Fillips, containing this statement.}. Kinematics
can't save the vacuum from decaying.

There is another way to describe this decay. We can imagine the patches of
constant curvature embedded in the flat Minkowski space. One can define the
standard Minkowski energy , which is always positive but is not strictly
conserved, due to the time dependence of the patches (which undergo
accelerated expansion). It has a typical "uncertainty" $(\Delta E)^{2}\sim R,$
where $R$ is the curvature of the de Sitter patches \cite{talk}, see also
\cite{akh}.  The positively curved patches of space-time behave as an
inversely populated medium, able to provide energy, generate particles from
nothing or amplify the propagating waves. This \textit{" cosmic laser" }will
continue to operate until all the curved patches are flattened. We can say
that the domains with cosmic acceleration create particles and the
gravitational attraction of these particles slows down the acceleration.

On a fixed background the density of the generated particles may or may not
blow up since they keep being produced and at the same time decay. The blowup
is crucial for our project, since in the case of finite density the back
reaction would be of the order $R/M_{pl}^{2}$ and thus negligible. The
screening of the cosmological constant may occurs due to an avalanche created
by the stimulated radiation. We have to develop the "de Sitter kinetics" to
describe this process.

It is important to realize that we must discuss interacting theories. At the
same time " particle creation" in the free theories has been discussed in the
numerous papers, usually by means of the Bogolyubov transformation. There is
nothing wrong with these discussions, except that they preserve the de Sitter
symmetry and thus can' t lead to the dynamical screening of the cosmological
constant. This effect ( due to the unbroken symmetry) just gives an infrared-
finite renormalization of this constant (which must remain a constant due to
the symmetry). At the same time the interaction may lead to the spontaneous
breakdown of the symmetry, making the curvature time-dependent.

Yet another view of the problem arises from the functional integral over the
metrices, $Z=\int Dge^{iS(g)}.$ In the semi-classical limit this integral is
dominated by various backgrounds. As we will see, the effective action for the
de Sitter space has a non-zero imaginary part, proportional to the infinite
volume, and thus is exponentially suppressed in the functional integral. This
conforms the definition in \cite{pol08} requiring the "eternal" manifolds to
have ( among other things ) real effective action. However, in the present
article we concentrate on the properties of the fixed backgrounds.

Our project for solving the cosmological constant problem can now be described
as follows. We start with some "bare" cosmological constant with its repulsive
nature , which causes constant acceleration of the universe. The "cosmic
laser" generates more and more particles. Their energy-momentum, including
mutual attraction, should slow down the acceleration and may eliminate it in
the infinite future. In this paper we will study only the first part of the
problem, leaving the back reaction for another publication.

\section{ The choice and use of the propagators}

To address quantitatively the problem of stability we have to know the
propagators and Feynman's rules. In quantum gravity this is not a trivial
problem. I discussed it in the previous paper \cite{pol08} , There are several
possible choices for the massive propagators. One can start from a sphere on
which the propagator is unique and then analytically continue it to the de
Sitter space. This procedure was first suggested by Chernikov and Tagirov
\cite{chern67} and has been discussed in the innumerable papers. The
corresponding state is the so called Bunch- Davies vacuum. This propagator has
the form%
\begin{equation}
G(n,n^{\prime})=\frac{1}{\sin\pi\nu}C_{\nu}^{d/2}(-nn^{\prime})
\end{equation}
Here the dimension of the dS space is $D=$ $d+1,$ The space is spanned by the
vectors $n$ satisfying the relation $n^{2}\equiv\overrightarrow{n}^{2}%
-n_{0}^{2}=1$ , $C$ is the Gegenbauer function ( which is the Legendre
function $P$ for $D=2$ and its first derivative for $D=4.$ We also have
$\nu=-d/2+i\mu,$ $\mu=\sqrt{M^{2}-\frac{d^{2}}{4}}$ , $M$ being the mass of
the particle; the Hubble constant is set to one, $H=1$. In this paper we will
concentrate on the "heavy " particles, $M>d/2,$ leaving an interesting case of
light and massless particles for another article. From the mathematical point
of view, the heavy particles belong to the main series of representations,
while the light ones correspond to the complementary series.

As was noticed in \cite{pol08} , the asymptotic behavior of this propagator at
large time-like separation has the structure
\begin{equation}
G\sim z^{-\frac{d}{2}}(A(-z)^{i\mu}+A^{\ast}(-z)^{i\mu})\sim e^{-\frac{d}{2}%
L}(Ae^{-\pi\mu}e^{i\mu L}+A^{\ast}e^{\pi\mu}e^{-i\mu L})
\end{equation}
where $L$ is the geodesic distance, $z=$ $(nn^{\prime})=\cosh L.$ This
asymptotic behavior means that this propagator can't be obtained by sum over
paths (which would always give a single exponential). This in turn means that
it is inconsistent to use this propagator in the standard perturbation theory
. Formally this is seen from the fact that almost all Feynman's diagrams are
divergent in the infrared due to the non-oscillatory terms in the product of
the Green's functions. Physically it is instructive to compare this behavior
with the one in the Minkowski space. The presence of the two exponentials
simply indicates that the system can experience transitions both increasing
and decreasing energy. This is the "inverse population" we mentioned above.
According to the above formula, the ratio between emission and absorption is
equal to $e^{-2\pi\mu}$ which looks like the effect of the Gibbons -Hawking
temperature. However, it is important to remember that we are working
exclusively with pure states and not a thermal mix. The meaning of the two
exponentials is that the corresponding state can transfer energy to an
external particle interacting with it (e.g. to the Unruh detector) without,
however, destroying the purity of the states.

We don't want to say that this propagator is meaningless. It can be
interpreted as an in/in propagator and one must use it in the Schwinger-
Keldysh perturbation theory . A complementary approach to the theory is based
on the in/out matrix elements. In this case we have the standard Feynman
rules, but the propagator $G_{F}(n,n^{\prime})$ is different from the above
and contains only one exponential. It is expressed in terms of the Legendre
$Q-$ function and (in contrast with the above propagator) can be represented
as a sum over trajectories lying on the hyperboloid. In this case the inverse
population manifests itself as an imaginary part of this Green function,
indicating an instability.

Before moving further, let us explain the meaning of the "in" and "out"
states. It is convenient for this purpose to use the Poincare coordinates
$ds^{2}=\rho(\tau)(dx^{2}-d\tau^{2})$ . Let us imagine that in the infinite
past we started with the flat space and then the expansion begins
adiabatically and we end up with the de Sitter space . After that we can stop
the expansion if necessary. A convenient model for the first stage is the
geodesically complete metric $\rho(\tau)=1+\frac{L^{2}}{\tau^{2}}$ where $L$
is the radius of the de Sitter. If we consider the Klein-Gordon equation in
this metric, the positive frequency solution ( defined by its asymptotic at
$\tau\rightarrow\infty$ ) is just $\varphi\sim\tau^{\frac{d}{2}}H_{i\mu}%
^{(1)}(\sqrt{p^{2}+m^{2}}\tau$ ), where $\mu=\sqrt{m^{2}L^{2}-\frac{d^{2}}{4}%
}$ , and $p$ is the momentum. This solution becomes the standard de Sitter
solution at small enough $\tau$ which gives us the "$in"$ vacuum. In order to
define the " $out"$ vacuum, we have to flatten the space near the future
infinity ( $\tau\rightarrow0).$ The de Sitter solution which matches
adiabatically with the plane wave must behave as $\tau^{i\mu}$ and thus is
proportional to the Bessel function $J_{i\mu}(p\tau).$ The above construction
describes the expanding universe ; by time reversal we can get a contracting
universe as well.

It has been noticed many times that this change of the basis from $"in"$ to "
$out"$ wave functions implies particle production. However, this effect is not
sufficient for an interesting back reaction because the de Sitter symmetry is preserved.

This becomes quite obvious if we look at some physical expectation value. For
instance, the quantity $\langle in|\varphi^{2}|in\rangle$ ,which can be easily
evaluated using the above solutions, grows a little in the transition but then
stays constant due to the de Sitter symmetry of the above propagator. If we
look at the expectation of the energy - momentum tensor, we see the same
picture - due to the symmetry it gives some unimportant renormalization of the
cosmological constant. The effect we are after comes from interaction and is
discussed below. This should be contrasted with Schwinger's pair production,
say in $1+1$ dimensions. The electric field in this case is a Lorentz
pseudoscalar, but the $"in"$ state breaks the Lorentz symmetry. As a result we
get a non-zero current generated by the created particles, producing screening
effect \cite{cooper}.

As we have said, in order to define unambiguously the " particle production"
,as well as "in" and "out" states, we have to join the de Sitter space (or any
other curved space for that matter) with the asymptotically flat spaces in the
past and in the future. This is analogous to the well known fact that we need
such asymptotic behavior in order to define energy and momentum in general. In
both cases we need a safe haven of the flat space to describe these objects.
We gave one example of such a metric in the above discussion. More generally,
for the metric $ds^{2}=dt^{2}-a^{2}(t)dx^{2}$ , we have the WKB wave function
of the form
\begin{equation}
\varphi\sim\exp-i\int^{t}\sqrt{M^{2}+p^{2}/a^{2}(t)}dt
\end{equation}
We can assume that $a(t)=e^{t}$ if $t_{i}$ $\ll t\ll t_{f}$ and tends to the
constants outside this interval. This behavior describes the expanding
universe. The plane waves of the Minkowski part of space- time match the above
wave functions $H_{i\mu}^{(1)}(pe^{-t}).$ In the WKB regime , defined by the
condition $\frac{d}{dt}(a/p)\ll1$ or $p\gg\frac{da}{dt},$ the reflection
coefficients are exponentially small and there is negligible particle
production at the matching stage. After the exponential period we get two
terms $e^{\pm i\mu t}$ from the asymptotic of the Hankel function, which means
that at the future infinity we have particles. Schematically, we have
\begin{equation}
e^{ipt}\Rightarrow e^{ipe^{-t}}\Rightarrow H_{i\mu}^{(1)}(pe^{-t})\Rightarrow
Ae^{i\mu t}+Be^{-i\mu t}%
\end{equation}
In the global coordinates the corresponding mode is $(\cosh t)^{\frac{1}{2}%
}P_{-\frac{1}{2}+i\mu}^{p}(i\sinh t)\approx H_{i\mu}^{(1)}(pe^{-t})$ as
$t\rightarrow\infty.$

Notice that the above considerations are specific for the expanding universe.
For the contracting case we would get a different set of modes and
propagators. Consider for instance a theory on a sphere of the radius $a(t)$
which tends to a constant as $t\rightarrow\pm\infty$ and behaves as
$a(t)=\cosh t$ in the intermediate region. It describes contraction followed
by expansion. In this case the we have to use the modes $(\cosh t)^{\frac
{1}{2}}Q_{-\frac{1}{2}+i\mu}^{p}(i\sinh t)\approx J_{i\mu}(pe^{-t}).$

It is very well known that the de Sitter space can be presented in the
Friedman form in three equivalent ways, having positive, zero or negative
spatial curvature. When we place a de Sitter patch in the Minkowski space, we
see that these ways are not equivalent anymore - the Minkowski vacuum matches
with the different de Sitter vacua. Let us notice in passing, that the often
used argument that the de Sitter modes must become the Minkowskian modes at
large momenta is not correct in general. The state of the de Sitter space
depends on the way one matches it with the Minkowski at the boundary. In the
bulk we can easily find the state with the non-zero occupation numbers (from
the Minkowski point of view).

Another important feature of the standard (P) propagator is the following. In
Minkowski space the Wightman functions can be obtained from the Euclidean
correlators by taking time to the lower half -plane where the positive
frequency solutions are bounded. Analogously, in the de Sitter case we can
analytically continue the propagator to the Euclidean half-sphere and the
corresponding eigen-modes will have no singularity on the south pole. This is
straightforward in the global coordinates which are obtained from the sphere
by the change $\vartheta\rightarrow it+\frac{\pi}{2},$ where $\vartheta$ is
the polar angle on a sphere and $t$ is the global time on the dS . Choosing
the eigenmodes to be the Legendre $P$ functions we get the desired
analyticity. More than that, in the Minkowsky space we define the positive
frequency solutions as the one which can be continued to the lower half-plane
of complex time. Analogously, the eigenmodes defining the propagator ( 1) can
be defined by their regularity in the southern hemisphere. One is tempted to
guess that, like in the Minkowski space, it may be possible to calculate (in
the interacting theory) the correlators on a sphere and then go to the de
Sitter space by analytic continuation. However, this guess is $^{{}}$incorrect.

\section{Centaurus and hyperboloids}

After the qualitative discussion above ( some of which perhaps is not new),
let us specify the modes, we will be working with. Let us again consider free
fields and expand them as%
\begin{equation}
\varphi(n)=(\cosh t)^{-\frac{d}{2}}\sum_{p}a_{p}f_{p}^{\ast}e^{ip\alpha}%
+a_{p}^{+}f_{p}e^{-ip\alpha}%
\end{equation}
where $a,$ $a^{+}$ are the usual creation and annihilation operators and the
eigenmodes $f$ of the Klein -Gordon operator satisfy%
\begin{equation}
\lbrack\partial_{t}^{2}+(\mu^{2}+\frac{(p+\frac{d-1}{2})^{2}-1/4}{\cosh^{2}%
t})]f_{p}=0
\end{equation}
To simplify notations, we consider the $d=1$ case where the metric is
$ds^{2}=-dt^{2}+\cosh^{2}td\alpha^{2^{{}}}$ ; general dimensionality requires
slight cosmetic changes in the formulae (in particular, the replacement of the
exponentials in ( 5) by the d-dimensional spherical functions). This metric is
obtained from the standard metric on a sphere by taking the polar angle
$\vartheta$ =$\pi/2+it.$ For a free field the choice of the vacuum is
synonymous to the choice of the eigenmodes $f.$ The standard propagator is
obtained if we take the modes proportional to $P_{l}^{p}(\cos\vartheta
)=P_{l}^{p}(i\sinh t),$ with $l=-1/2+i\mu.$ They are distinguished by being
non-singular at $\vartheta=0$ (on a sphere we would also require analyticity
at $\vartheta=\pi$ which would quantize $l).$ We denote this vacuum as
$|E\rangle.$ A different vacuum is obtained by choosing $f$ to be a Jost
function i.e. behaving as $e^{i\mu t}$ at the past or the future infinities,
as was discussed above. They are described by $Q_{l}^{p}(\cos\vartheta).$ We
will call this state $|J\rangle.$ From the above discussion it is clear that
this state describes contracting and then expanding case,

When we introduce the interaction we have to make another choice, the choice
of the manifold. Namely, we have to decide how to perform the integrations in
Feynman's diagrams. For the $E-$ state we can either assume that the manifold
is a full hyperboloid, in which case we have to integrate over $t$ from
$-\infty$ to $+\infty,$ or the well known manifold consisting of the half-
sphere with the Euclidean signature glued to the half hyperboloid with the
Minkowskian signature. In this case we have to choose the complex contour of
integration over $t$ , consisting of two segments $[i\pi/2,0]\cup
\lbrack0,+\infty].$

The physical reason for such a a centaur-like manifold is that it represents a
tunneling creation of the de Sitter universe from nothing \cite{vile}
\cite{HH}. It is technically analogous to the treatment of tunneling and
subsequent propagation of a quantum particle. In this case we take a classical
solution in imaginary time for the under-the barrier period and join it ( at
the stopping point) to the solution in real time when the barrier is already
crossed. All corrections to the semi-classical approximation in this case are
given by Feynman's diagrams in which the integration goes over the complex
contour in time.

The full hyperboloid picture assumes that the universe simply exists ,while in
the "centaur" picture it is created from nothing. Since the present author
doesn't dare to make such choices, we will discuss both possibilities.

\section{Obstruction to Wick's rotation and interaction instability}

In the previous work and in the above discussion we studied the free particles
instabilities. We visualized creation of the free particles which are then
taken apart by the expanding geometry. This is a precise analogue of
Schwinger's pair creation, in which case we have an electric field instead of
geometry . As we already mentioned, such a process doesn't create large enough
back reaction, since the in/in matrix elements are restricted by the de Sitter
symmetry. Things are different in the case of interaction.

Let us consider first the massive $\lambda\varphi^{4}$ theory in the global de
Sitter space. Let us assume that the system was in the Euclidean vacuum
described above. There is a Fock space built on the top of this vacuum. The
interaction, when applied, creates excitations. In order to study them, we
must examine the matrix elements of the $S=T\exp(-i\lambda\int\varphi^{4}dn).$
Here the integration goes over the complete de Sitter space. The chronological
ordering must be chosen so as to preserve the de Sitter symmetry. The simplest
choice is simply to order along the $n_{0}$ - direction. Indeed, the quantity
$sgn(n_{0}-n_{0}^{\prime})$ is invariant, providing that $(n-n^{\prime}%
)^{2}=2(1-nn^{\prime})\leq0$ (timelike). This is precisely where the
propagator (1 ) has a cut. So, just as in the Minkowsky space, the choice of
the approach to the cut defines the ordering of the operators and is invariant
under the symmetry group (for the space-like separations the order doesn't matter).

The effect we are after appears in the second order in $\lambda.$ The general
expression for the vacuum action has the form%
\begin{equation}
iW\equiv\log\langle0|S|0\rangle\sim(vol)\lambda^{2}\int dn^{\prime}%
G^{4}(nn^{\prime}-i0)
\end{equation}
Here the $(vol)$ is the total space-time volume of the de Sitter space, the
points of which are defined by the unit vector $n$ , satisfying the relation
$n^{2}\equiv\overrightarrow{n}^{2}-n_{0}^{2}=1.$ The integration is performed
with respect to the invariant measure, $dn=d\overrightarrow{n}dn_{0}%
\delta(n^{2}-1).$ The propagators in this formula depend on the physical
assumptions. As we argued above, in the present setting we must use ( ) which
are obtained by the analytic continuation of the unique propagator on the
sphere (related to the dS space by the "Wick rotation" $n_{0}\rightarrow
in_{4}$ ). If we could perform the Wick rotation in the above integral, the
effective action would be real and there would be no interesting effects, just
the shift of the vacuum energy. However, unlike the Minkowski space, the dS
space doesn't allow such a rotation, as we are going to demonstrate. We are
looking for the\emph{\ imaginary contribution to the effective action}.

In order to do the above integral it is convenient to take the vector $n$
having only one non-zero component, $n_{1}=1$ (due to the symmetry the
integral in (7 ) doesn't depend on $n$ ). The integral takes the form%
\[
i\Lambda\sim\lambda^{2}\int_{-\infty}^{\infty}dn_{1}G^{4}(n_{1}-i0)\int
dn_{0}d^{D-2}n_{i}\delta(n_{0}^{2}+1-n_{1}^{2}-n_{i}^{2})
\]
where $i$ $=2,...D$ and $D$ is the dimension of the dS and the action density
is introduced, $W=\Lambda\cdot(vol)$. Integrating out the $\delta- $ function
gives%
\begin{equation}
i\Lambda\sim\lambda^{2}\int_{-\infty}^{\infty}dn_{1}G^{4}(n_{1}-i0)\int
dn_{0}(n_{0}^{2}+1-n_{1}^{2})_{+}^{\frac{D-3}{2}}%
\end{equation}
where the subscript $+$ means that the expression in the brackets must be
positive. For the physical dimensions the integral over $n_{0}$ has a power-
like divergence. This divergence is not dangerous. If we introduce a cut-off
for $n_{0}$ , coming from the adiabatic turn-off of the interaction, the
leading divergence will be proportional to the $\int dn_{1}G^{4}(n_{1}-i0$
$)=0,$ since the integrand has no singularities in the lower half-plane. As a
result, we can understand the integral in (8 ) in the sense of analytic
continuation in $D,$ and express it in terms of the $\Gamma-$ functions. The
final answer is non-trivial and depends on whether $D$ is even or odd. Let us
write it in the cases $D=3$ and $D=4.$ For $D=4$ we have
\begin{equation}
i\Lambda\sim\int_{-\infty}^{\infty}dn_{1}(n_{1}^{2}-1)\log|n_{1}^{2}%
-1|G^{4}(n_{1}-i0)
\end{equation}
while for $D=3$
\begin{equation}
i\Lambda\sim(\int_{1}^{\infty}dn_{1}+\int_{-\infty}^{-1}dn_{1})\sqrt{n_{1}%
^{2}-1}G^{4}(n_{1}-i0)
\end{equation}
We now use these formulae to calculate $\operatorname{Im}W.$ There are useful
sum rules satisfied by these integrals%
\begin{align*}
&  \int_{-\infty}^{\infty}dn_{1}(n_{1}^{2}-1)\log((n_{1}^{{}}-i0)^{2}%
-1)G^{4}(n_{1}-i0)\\
&  =\int_{-\infty}^{\infty}dn_{1}(n_{1}^{2}-1)\log|n_{1}^{2}-1|G^{4}%
(n_{1}-i0)+i\pi\int_{-\infty}^{\infty}dn_{1}(n_{1}^{2}-1)_{+}sgn(n_{1}%
)G^{4}(n_{1}-i0)\\
&  =0
\end{align*}
together with $\int_{-\infty}^{\infty}dn_{1}(n_{1}^{2}-1)G^{4}(n_{1}-i0)=0$
for $D=4$ and analogous formulae for $D=3$ , following from the analyticity at
$\operatorname{Im}n_{1}<0.$%
\begin{equation}
\int_{-\infty}^{\infty}dn_{1}\sqrt{(n_{1}-i0)^{2}-1}G^{4}(n_{1}-i0)=I_{+}%
-I_{-}+i\int_{-1}^{1}(...)=0
\end{equation}
where $I_{+}=\int_{1}^{\infty}$ $(...)$ , $I_{-}=\int_{-\infty}^{-1}(...)$ and
the integrands are everywhere the same. Clearly the similar formulae hold in
any dimension, with the logarithms in the even ones and the square roots in
the odd .

If we combine these results with the fact that the function $G(n_{1})$ is real
for $n_{1}\leq1$ we come to a rather surprising conclusion. For $D=3$ (or for
any odd dimensions) we have
\begin{equation}
iW\sim I_{+}+I_{-}=2I_{-}+iI_{0}%
\end{equation}
where $I_{0}=\int_{-1}^{1}dn_{1}\sqrt{1-n_{1}^{2}}G^{4}(n_{1}).$ This integral
is precisely the action we get on a sphere. It gives the real part of the
action on the hyperboloid. The integral $I_{-}$ $,$ which is also real,
represents the obstruction to the Wick rotation and gives the imaginary part
of the effective action.
\begin{equation}
\operatorname{Im}\Lambda\sim\lambda^{2}\int_{-\infty}^{-1}dn_{1}G^{4}(n_{1})
\end{equation}
For the even dimensions the answer is different. From the above sum rules, at
$D=4$ we get%
\begin{equation}
i\Lambda\sim i\pi J_{0}+2i\pi J_{-}%
\end{equation}
where once again $J_{0}=\int_{-1}^{1}(n_{1}^{2}-1)G^{4}(n_{1})dn_{1}$ and
$J_{-}=\int_{-\infty}^{-1}(...).$ The second term represents the obstruction
to Wick's rotation but for even dimensions this obstruction gives real
effective action \footnote{I was informed by U. Moschella, that this result
was known to him and his collaborators in the different framework}. As we will
see below, this doesn't mean that the theory is stable. It is interesting to
note that for the non-interacting case the situation with the effective action
is the opposite. Namely, for odd dimension we get a reflectionless potential
in the Klein Gordon equation which means that there are no creation of
Bogolyubov "particles", while in the even dimension such "particles" are
created. It would be nice to have a deeper understanding of this fact.

There is another subtle point which must be stressed. We assumed that we
switch the interaction adiabatically. However, due to the " adiabatic
catastrophe" \cite{pol08} in this process we create additional particles. The
origin of this phenomenon lies in the non-positivity of energy (this contrasts
with the Minkowsky space where the slow changing fields don't have energy to
create particles). In the above calculations we concentrated on the bulk of
the de Sitter space, excluding the particles created in the infinite past when
the interaction was turning on. Technically this can be seen from the formula%
\begin{equation}
iW\sim\int dndn^{\prime}\lambda(\epsilon n_{0})\lambda(\epsilon n_{0}^{\prime
})G^{4}(nn^{\prime})
\end{equation}
where we introduced adiabatic time dependence ( $\epsilon\rightarrow0)$ of the
coupling constant. The above calculations ( and cancellations) refer to the
part of this integral where $n_{0},$ $n_{0}^{\prime}$ $\sim1.$ At the same
time the domain $n_{0},n_{0}^{\prime}\sim1/\epsilon$ will also contribute
(unlike the Minkowski case). These are the excitations created in the
asymptotic past and future by the switching of the interaction. This part of
the integral depends on the concrete function $\lambda$ and is hardly interesting.

The case of Centaurus is very similar. In this case the variable $n_{0}$ runs
along the contour in the complex plane, $n_{0}\subset\lbrack0,\infty
]\cup\lbrack-i,0].$ The above integrals are again real in the even dimensions
and have imaginary part in the odd case.

In the case of the Jost vacua, defined by the $Q$ eigenmodes, which are the
Jost functions for the equation (6 ) , we get the non-zero imaginary part in
all dimension. This is because the $Q$ functions have an antipodal
singularity. In a certain sense this makes them less stable than the Euclidean vacuum.

\section{Vacuum vs. particle instability}

It is instructive to have another interpretation of these results. If we plug
the expansion (5 ) into the $S$ matrix we get the following production
amplitude%
\begin{align}
S|0\rangle &  \sim i\lambda\sum_{p}\int f_{p_{1}}...f_{p_{4}}dn|p_{1}%
...p_{4}\rangle\\
&  =i\lambda\sum_{p}\int P_{l}^{p_{1}}(i\sinh t)...P_{l}^{p_{4}}(i\sinh
t)\cosh tdt(a_{p_{1}}^{+}...a_{p_{4}}^{+})|0\rangle
\end{align}
here we again look at the 2d case to simplify notations. We see that the
production amplitude is given by
\begin{equation}
\Gamma(p_{i})=\int_{-i\infty}^{i\infty}dxP_{l}^{p_{1}}(x)...P_{l}^{p_{4}}(x)
\end{equation}
here $l=-1/2+i\mu$ and the magnetic quantum numbers satisfy the conservation
law $\sum p_{i}=0.$ In D- dimensions we have to replace as before the Legendre
functions with the Gegenbauer ones and to change the measure to $(1-x^{2}%
)^{D/2-1}dx.$ The Legendre functions have one branch point at $x=-1,$ and a
kinematic singularity of the form $(1-x)^{|p|/2}.$ Due to the above
conservation law, the $\sum|p_{i}|$ is an even number.Hence the contour of
integration can be displaced to infinity, making the integral equal to zero,
$\Gamma(p_{i})=0$ for $D=2n$. A useful formula to check these facts for
general $D$ is $f_{p}\sim(z^{2}-1)^{-\frac{d-1}{4}}P_{-\frac{1}{2}+i\mu
}^{-(p+\frac{d-1}{2})}(z)\sim(z^{2}-1)^{\frac{p}{2}}F(z)$ where $F(z)$ has
only one branch point at $z=-1$ ; here $d=D-1.$ It is easy to check that the
sum of the angular momenta $p$ is always even (this follows from the spatial
reflection symmetry). In the odd dimensions there is another branch point at
$x=1$ coming from the measure and we get the non-zero production. As was
explained in \cite{pol08}, the zero of the above overlap integral means the
absence of the "spider diagrams " in the Schwinger -Keldysh formalism and is a
necessary condition for the vacuum stability ( in the flat space this integral
is zero due to the energy conservation). These statements must also have a
group-theoretic interpretation since the overlap integrals are related to the
Clebsch-Gordan coefficients. It is also curious to notice that without
interaction the stability situation is the opposite - the Jost vacuum is
stable in odd dimension since the equation (6 ) has a KdV soliton as the potential.

However, when we look at the higher loops diagrams in even dimensions, we get
the structures like $\int dt_{1}..dt_{4}F(t_{1}...t_{4})f(t_{1})...f(t_{4})$ ,
where $F$ is the four point amputated amplitude. Generally speaking, this
amplitude is expected to have the branch points both at $z=1$ and $z=-1,$
leading to vacuum decay. However I haven't checked this by direct calculation.
It might happen that some hidden discrete symmetry would invalidate this
conclusion. Deeper understanding of this subtle question is desirable, but in
the following we will assume that the production amplitudes are indeed non-zero.

Stability of the vacuum, however, is not all. We have to study the one
particle stability as well. In this case we have
\begin{equation}
S|p\rangle\sim i\lambda\sum_{p_{k}}\int f_{p}^{\ast}f_{p_{1}}...f_{p_{3}%
}dn|p_{1}...p_{3}\rangle
\end{equation}
This time the overlap is non-zero, since the first factor, $f^{\ast},$ has a
left cut in the $z=i\sinh t$ complex plane. In the even- dimensional case the
production amplitude can be written as an integral along this cut. As a
consequence, the particles decay. This instability has been recognized in the
early days of the subject \cite{nacht67}, \cite{mirhv} and recently discussed
in \cite{vega} \cite{bros}, \cite{vol09}. We will move a step further and
study the development of a state with the prescribed occupation numbers. Due
to the above instability, these occupation numbers change in time and our task
is to find the kinetic equation describing this process.

A convenient way to do this is to begin with the identity%
\begin{equation}
\frac{1}{T}\langle N_{out}-N_{in}\rangle=\frac{1}{T}\sum_{M}(M-N)|\langle
M|S|N\rangle|^{2}%
\end{equation}
where $M,N$ are the occupation numbers for the various modes and $S$ is the
scattering matrix ,defined for the large but finite time interval $T.$ If we
take a generic initial state, the right hand side of this equation will
produce infrared divergences as $T\rightarrow\infty.$ This simply means that
the interaction changes the original occupation numbers with the non-zero
rate. If the interaction is small, these numbers change slowly and we can
replace the left hand side by $\frac{dN}{dT}.$ To calculate the right hand
side we use the perturbation theory and the above formulae. The result is the
sum of several terms in which some of the occupation numbers are changed by
$\pm1.$ For example the formula (16 ) corresponds to
\begin{equation}
(S-1)|\{N_{p}\}\rangle=\sum_{p_{1}...p_{4}}\Gamma(p_{i})\sqrt{(1+N_{p_{1}%
})...(1+N_{p_{4}})}|...N_{p_{1}}+1...N_{p_{4}}+1...\rangle
\end{equation}
where $\Gamma$ is the overlap integral discussed above, and we added the
obvious Bose factors. There are also similar terms in which the corresponding
occupation numbers are reduced. Since the identity does not contribute to the
right hand side of ( 22) we obtain from this formula the kinetic equation%
\begin{equation}
(\frac{dN_{p}}{dT})^{(1)}=(\frac{1}{T}\sum_{p_{i}}|\Gamma|^{2}\{(1+N_{p}%
)(1+N_{p_{1}})...(1+N_{p_{3}})-N_{p}...N_{p_{3}}\}+...)_{T\rightarrow\infty}%
\end{equation}
The terms displayed in this formula come from the four $a^{+}$ or four $a$
operators acting on the state. The terms indicated by the dots are those from
$a^{+}a^{3},$ ($a^{+})^{2}a^{2}$ and their conjugate, which can be easily
written explicitly. For example the first type of terms have the form%
\begin{equation}
(\frac{dN_{p}}{dT})^{(2)}\sim(\frac{1}{T}\sum|B|^{2}\{(1+N_{p})N_{p_{1}%
}...N_{p_{3}}-N_{p}(1+N_{p_{1}})...(1+N_{p_{3}})\}+...)_{T\rightarrow\infty}%
\end{equation}
The negative sign appears, according to (20 ), when the occupation number is
reduced (from the terms like $a^{4}$ and $a(a^{+})^{3}$ ). The overlap
amplitude $B$ (different from $\Gamma$ ) is the one entering ( 19). The
displayed term describes the process $(p)\leftrightarrow(p_{1}p_{2}p_{3})$ (by
this notation we mean that a particle with momentum $(p)$ decays onto three
particles together with the reversed process). As is well known \cite{LL} ,
one has to add to the collision integrals the terms describing $(p_{1}%
)\leftrightarrow(pp_{2}p_{3})$ also coming from the interaction of the type
$a(a^{+})^{3}+c.c.$ and the terms $(pp_{1})\leftrightarrow(p_{2}p_{3})$ from
$a^{2}(a^{+})^{2}$ \footnote{In the first version of this paper these terms
were lost, which led to the wrong sign in the eq, (27 ). I thank D. Krotov for
finding this error. The detailed derivations will be discussed elsewhere.} .
The general rule \cite{LL} is that we must write the factor $N$  for each
decayed particle and $1+N$ for the created one. The corresponding product of
the Bose factors enters with a plus if it increases the value of $N_{p}$ and
with a minus otherwise.

This kinetic equation is unusual in two respects. First, the $\Gamma$ - term,
which is strictly positive, normally doesn't appear in statistical physics due
to energy conservation. Second, in the standard case we have Boltzmann's
H-theorem, ensuring that there exist the equilibrium with maximal entropy.
This is not the case here (positivity of energy and its conservation are
crucial for the H-theorem). We also have to evaluate the large time limits in
these equations. The time dependence of the collision terms in ( 22)and (23)
arises due to their logarithmic divergence at large $p$ and the fact that due
to the expansion the invariant cut-off exponentially depends on time.

In order to see it let us find the asymptotic of these expressions at large
$p.$ For that we shall use the (already mentioned) fact that in this limit the
Legendre functions in the integrand can be replaced by the Bessel or Hankel
functions. For example, $(\cosh t)^{1/2}P_{\frac{-1}{2}+i\mu}^{p}(i\sinh
t)\rightarrow H_{i\mu}^{(1)}(pe^{-t})$ if $p\gg1$ and $t>0.$ Using this and
other similar formulae, it is easy to find the asymptotic behavior of the
amplitudes $\Gamma$ and $B.$ Let us restrict ourselves with the scalar fields
with $\varphi^{N}$ interaction in the space ($dS)_{d+1}$ . As we discussed
above, in this case $d+1$ must be odd, otherwise the amplitudes are equal to
zero. We get%
\begin{equation}
\Gamma_{N}(p_{i})=C_{N}\int_{0}^{\infty}\frac{d\tau}{\tau}\tau^{\frac
{d(N-2)}{2}}H_{i\mu}^{(1)}(|p_{1}|\tau)...H_{i\mu}^{(1)}(|p_{N}|\tau)
\end{equation}
where $\tau=e^{-t}.$ From this it is obvious that $\Gamma_{N}$ scales as
$p^{-\frac{d(N-2)}{2}}.$ The other important feature of this quantity is that
in the limit $p_{1}\rightarrow0$ it behaves as $\Gamma\rightarrow Fp_{1}%
^{i\mu}+c.c$ , where $F$ is a function of the other momenta. As a result
$|\Gamma|^{2}$ $\rightarrow|F|^{2}+...$ in this limit. Substituting this
result into the collision integral, we find a logarithmic divergence, provided
that the occupation numbers are bounded at infinite $p.$ This follows from the
fact that, due to the momentum conservation, we have precisely $N-2$
integrations over $d-$ dimensional momenta. The invariant cut-off $\Lambda$
restricts  the comoving momenta, $p\leq$ $p_{\max}=\Lambda e^{T}.$ In this way
we obtain the expected linear time dependence. We find that the collision
integral is a coefficient in front of the logarithm of the cut-off, making it
look  \emph{like an infrared analogue of the  beta function}.

Using this observation it is easy to rewrite the collision integrals in terms
of physical momenta, $f_{k}$. For that let us insert into (22 ) and ( 23) the
identity $1=\int dTe^{T}$ $\delta($ $|p|+\sum|p_{k}|-e^{T})$ and rescale all
the momenta by $p_{k}\Rightarrow e^{T}f_{k}$ . Using the above scale
invariance we obtain the kinetic equation for the occupation numbers in the
physical momentum space, $n(f,T)=N(fe^{T},T)$%
\begin{align}
&  \frac{\partial n}{\partial T}-f\frac{\partial n}{\partial f}\\
&  =\int%
%TCIMACRO{\dprod }%
%BeginExpansion
{\displaystyle\prod}
%EndExpansion
df_{k}\delta(f+\sum f_{k})\delta(|f|+\sum|f_{k}|-1)|\Gamma(f_{k}%
)|^{2}\nonumber\\
&  \{(1+n(f))(1+n(f_{1}))...(1+n(f_{3}))-n(f)n(f_{1})...n(f_{3})\}\nonumber
\end{align}
and analogously for the $B$ term. The $-1$ in the argument of the second delta
function reflect the energy non-conservation discussed at the beginning, while
the second term of the LHS is just a red shift .

While we have not explored the solutions of this equations in full generality,
there is an interesting special solution which reveals the underlying physics.
Returning to ( 22) , let us assume that $N(p,T)=N(T)$ for $|p|\ll\Lambda
e^{T},$ thus representing a "\ \emph{Bose sea} " . Let us also remember that
the quantity $|\Gamma(p_{k},p)|^{2}$ in the limit $p_{k}\gg p$ doesn't depend
on $p$ , which makes our assumption self-consistent. The resulting kinetic
equation has the form%
\begin{equation}
\frac{dN}{dT}=a[(1+N)^{4}-N^{4}]+2b[(N(1+N)^{3}-N^{3}(1+N)]
\end{equation}
The (positive) constants $a$ and $b$ are the corresponding integrals of
$\Gamma$ and $B$ terms . The first one represents the vacuum decay and the
reverse process while the second is the particle decay and the reverse. If
$N\rightarrow0$ , only the first term survives. Its presence is simply another
indication that the vacuum as well as the de Sitter symmetry is unstable. If
$N\rightarrow\infty,$ we get%
\begin{equation}
\frac{dN}{dT}\simeq(4a+4b)N^{3}%
\end{equation}
We see that we get a finite time singularity, $N\sim(T_{\ast}-T)^{-\frac{1}%
{2}}.$ This is the explosive regime discussed  above. The back reaction in
this regime becomes large.

This behavior is easy to understand qualitatively - particles are generated
faster than they decay. It is clear that the back reaction in this case
becomes unbounded.

Let us explain briefly the general relation between the Boltzmann equation and
infrared divergences. Consider a correction the Green's function%
\begin{align}
G^{(1)}(n,n^{\prime}) &  =\int dn_{1}dn_{2}G(nn_{1})\Sigma(n_{1}n_{2}%
)G(n_{2}n^{\prime})\\
\Sigma(n_{1}n_{2}) &  =\lambda^{2}G^{3}(n_{1}n_{2})\nonumber
\end{align}
where $G$ -s are the standard propagator supplied with the Schwinger-Keldysh
$\pm$ indices (see \cite{pol08} ) and the summation over these indices is
understood. Because of the interference of the two exponentials, the integral
is logarithmically divergent in the region $(nn_{1,2})\gg(n_{1}n_{2})\sim1.$
As was explained in \cite{pol08} , this is a linear divergence in proper time.
In the Minkowski space or in condensed matter physics we get a similar linear
infrared divergence if we use the Green's functions with the states having
arbitrary occupation number. The meaning of this divergence is simple - the
interaction is changing the occupation number at constant rate, since they are
not in the equilibrium. The only exception is the set of the occupation
numbers (if it exists) for which the collision integral is zero. The role of
the Boltzmann equation is that it sums up the leading terms of the type
$(\lambda^{2}T)^{n}$ , which is good enough for small $\lambda.$ Incidentally,
this point of view provides a convenient approach for deriving the standard
Boltzmann equations from the Schwinger- Keldysh diagrams. The substantial
difference of our case is the absence of the H-theorem and thus the lack of
the guaranteed equilibrium state. Because of that the physical matrix
elements, like already mentioned $\langle in|\varphi^{2}(n)|in\rangle
=G^{(1)}(n,n)$ have the linear infrared divergence. It means, among other
things, that they are sensitive to the breakdown of the de Sitter symmetry in
the infinite past; this can be called the spontaneous breakdown of this symmetry.

We see that the above infrared divergence is related to the wrong (unstable)
choice of the initial state and not to the anomalously strong interaction. In
some sense it is analogous to the Bloch-Nordsieck divergence in QED. The
situation will be different in the case of light and massless particles. For
such particles $M^{2}\leq\frac{d^{2}}{4}$ (in the units of Hubble's constant)
and thus $\mu$ is purely imaginary, $\mu=i\lambda=i\sqrt{\frac{d^{2}}{4}%
-M^{2}}.$ Such particles correspond to the complementary representations of
the de Sitter group. This time the interaction becomes strong. This can be
seen from the overlap integral ( 26) which becomes divergent if $\frac
{d(N-2)}{2}<N\lambda.$ In this case we need new methods. Some possible
approaches, based on the summation of the leading terms, will be discussed
elsewhere. Another outstanding problem, which should be addressed, is the
effect of the back-reaction as well as the fluctuations of the background.
Although there are many unanswered questions and future surprises, I believe
that a small step made in this work is the step in the right direction.

I would like to thank T. Damour, J. Maldacena, U. Moschella, V. Mukhanov, D.
Podolsky, P. Steinhardt , G. Volovik and S. Weinberg for interesting and
helpful discussions. This work was partially supported by the NSF grant PHY-0756966.

\end{document}